\documentclass[prl,aps,twocolumn,superscriptaddress,showpacs,amsmath,amssymb,floatfix]{revtex4}

\usepackage{graphicx}

\newcommand{\be}{\begin{equation}}
\newcommand{\ee}{\end{equation}}

\begin{document}
\title{Formation of molecules near a Feshbach resonance in 
a 1D optical lattice}
\author{G. Orso}
\affiliation{Dipartimento di Fisica, Universit\`{a} di Trento and
BEC-INFM, 1-38050 Povo, Italy}

\author{L.P. Pitaevskii}
\affiliation{Dipartimento di Fisica, Universit\`{a} di Trento and
BEC-INFM, 1-38050 Povo, Italy}
\affiliation{Kapitza Institute for Physical Problems, 117334 Moscow, Russia}

\author{S. Stringari}
\affiliation{Dipartimento di Fisica, Universit\`{a} di Trento and
BEC-INFM, 1-38050 Povo, Italy}
\affiliation{\'{E}cole Normale Sup\'{e}rieure and Coll\`{e}ge de France, 
Laboratoire Kastler Brossel,
24 Rue Lhomond 75231 Paris, France}

\author{M. Wouters}
\affiliation{Dipartimento di Fisica, Universit\`{a} di Trento and
BEC-INFM, 1-38050 Povo, Italy}
\affiliation{TFVS, Universiteit Antwerpen, Universiteitsplein 1,
2610 Antwerpen, Belgium}

\begin{abstract}
We calculate the binding energy of two atoms interacting near a Feshbach
resonance in the presence of a 1D periodic potential. 
The critical value of the scattering length needed to
produce a molecule as well as the value of the molecular binding energy in
the unitarity limit of infinite scattering length are calculated as a
function of the intensity of the laser field generating the periodic
potential. The Bloch bandwidth  and the effective mass of molecules
are shown to depend strongly on the value of the scattering length
due to the correlated motion of the two atoms.
\end{abstract}
\maketitle

The availability of tunable periodic potentials of optical nature
(see, for example, Ref.\cite{varenna}) and the quantum 
manipulation of the interatomic forces via
Feshbach resonances \cite{ketterle} are opening new perspectives in 
the field of ultracold atomic gases.  
In the absence of the external potential molecules can be formed
near resonance only for positive values of the s-wave scattering length
$a$, the corresponding binding energy being given by the formula
$E_b=-\hbar^2/ma^2$ where $m$ is the mass of each atom. These molecules
have already been observed experimentally by several groups \cite{jila,mit,Innsbruck,ens}.
An interesting and very rich problem is the formation of 
molecules in an optical lattice
generated by laser fields. Recently, Fedichev et al. \cite{Fedichev_PRL92}
investigated the case of a 3D tight lattice and showed that
the bound state can be formed also for negative 
scattering lengths down to a critical value. Their approach is valid
provided that the binding energy is small compared to the width of the 
lowest Bloch band so that the molecular wavefunction extends 
over many lattice sites.

In this Letter we discuss the formation of molecules in a 1D optical
lattice for {\sl arbitrary} values of the scattering length and the laser 
intensity.
If the barriers separating two consecutive wells are
very high (high laser intensity) the system behaves like a series of 2D
discs which lie at the bottom of the wells and feel a harmonic potential in
the direction of the laser field.  In this asymptotic regime the two-body
problem is significantly simplified by the possibility of separating the
center of mass and the relative coordinates of the two interacting atoms 
\cite{gora}. When this condition is violated many interesting
questions can be addressed: what is the critical value of the scattering 
length required to ensure the formation of a molecule? 
What is the value of the
binding energy of the molecule in the so called unitarity limit where the
scattering length approaches an infinite value? 
How do the tunneling properties (bandwidth and effective mass) 
of molecules depend on the value of the scattering length?  
The purpose of this Letter is to provide an explicit answer to 
these questions.

Let us start by writing the 3D Schrodinger equation for two atoms interacting 
in the presence of a 1D periodic potential $V_{opt}(z)$:
\begin{eqnarray}
  \left(-\frac{\hbar^2}{m}\nabla^{2}-\frac{\hbar^2}{4m}\frac{\partial^2}{\partial Z^2}+V(Z,z) +g\delta(\mathbf{r})\frac{\partial }{\partial r} r\right)
\Psi=E\Psi,
\label{schrodinger}
\end{eqnarray}
where $V(Z,z)=V_{opt}(z_1)+V_{opt}(z_2)$ and 
we have introduced the center of mass and relative
coordinates  $Z=(z_1+z_2)/2$ and $\mathbf{r}=\mathbf r_1-\mathbf r_2$.

 In Eq.(\ref{schrodinger}) the interaction is modeled via 
a $s$-wave pseudopotential with coupling constant $g=4\pi \hbar^2 a/m$, 
where $a$ is the scattering length in vacuum. 
The solution of Eq.(\ref{schrodinger}) can be written, for $g\neq 0$, as  
\begin{equation}
  \Psi(\mathbf{r},Z)=\int dZ^{\prime} G_E(\mathbf{r},Z;\mathbf{0},Z^{\prime}) 
  g \frac{\partial }{\partial r^\prime} \left(r^\prime \Psi(\mathbf{r}^{\prime},Z^{\prime})\right)_{\mathbf{r}^\prime=0},
  \label{schrodinger2}
\end{equation}
where $G_E$ is the free $(g=0)$ Green's function associated to 
Eq.(\ref{schrodinger}). At short distance, the Green's function 
is determined only by the kinetic energy. Neglecting the external potential
and setting $E=0$ in Eq.(\ref{schrodinger}), we find 
$G_{E=0}(\mathbf{r},Z;\mathbf{0},Z^{\prime})=-(m/8\pi^2\hbar^2)[(Z-Z^\prime)^2+(r/2)^2]^{-1}$. Therefore
the Green's function has the following expansion
\begin{equation}
  G_E(\mathbf{r},Z;\mathbf{0},Z^{\prime})
 = -\frac{m}{4\pi \hbar^2 r}\delta\left( Z-Z^{\prime }\right)
+K_{E}\left(Z,Z^{\prime }\right)+O(r),  \label{defBC}
\end{equation}
where the regular kernel $K_E(Z,Z^{\prime })$ depends on the  
external potential and we used the representation of the delta function
$\delta(x)=\lim_{r\rightarrow 0}r/\pi(r^2+x^2)$. 
When inserted into Eq.(\ref{schrodinger2}), Eq.(\ref{defBC}) implies, 
for $r\rightarrow 0$, the boundary condition 
$\Psi(\mathbf{r},Z)\sim (1/r-1/a)f(Z)$, where $f(Z)$ is a function of
the center of mass position satisfying the integral equation
\begin{eqnarray}
\frac{1}{g}f\left( Z\right)  &=&\int dZ^{\prime }~K_{E}\left(
Z,Z^{\prime }\right) f\left( Z^{\prime }\right).   \label{int2}
\end{eqnarray}
If the center of mass and the relative motion decouple
  the function $f(Z)$ must be an eigenstate of the 
center of mass Hamiltonian independent
of the scattering length. By inserting it into Eq.(\ref{int2}),
the dependence on $Z$ factors out and we are left with an algebraic equation
yielding the energy as a function of the scattering length.
For instance, in the absence of the optical lattice the kernel is given by 
\be
K_E^{0}(Z,Z^{\prime })=\frac{m}{4\pi\hbar^2}\int \frac{dP}{2\pi}
  e^{i P (Z-Z^\prime)} \sqrt{\vert E\vert+\hbar^2 P^2/4m}.
  \label{GEfree}
\ee
The lowest energy solution of Eq.(\ref{int2}) corresponds to 
$f(Z)$ constant and yields $E=-\hbar^2/m a^2$.

For periodic potentials, the kernel has the property
 $K_E(Z,Z^\prime)=K_E(Z+j d,Z^{\prime }+j d)$
where $j$ is any integer and $d$ is the lattice spacing.
This means that the general solution of Eq.(\ref{int2}) 
has the Bloch form $f(Z)=e^{iQZ}f_{Q}(Z)$, where $f_{Q}(Z)$ is a 
periodic function, $Q$ being the quasi-momentum of the molecule.
In the following we shall restrict to the lowest bound state
given by Eq.(\ref{int2}). 
The energy as a function of the scattering length is then fixed by the 
condition that the lowest eigenvalue of the kernel $K_E(Z,Z')$ 
in Eq.(\ref{int2}) is equal to $1/g$.

Due to the singular term in Eq.(\ref{defBC}), the numerical evaluation
of the kernel $K_E(Z,Z^{\prime })$ from the Green's function $G_E$
is not trivial. The latter can be written in the general form
\begin{eqnarray}
 && G_E(\mathbf{r},Z;\mathbf{0},Z^{\prime}) =
  \sum_{n_1,n_2} \int\frac{d^{2}\mathbf{k}_{\perp }}{\left( 2\pi \right) ^{2}}
\int_{-q_B}^{q_B}\frac{dq_1}{2\pi}\frac{dq_2}{2\pi}
  e^{i\mathbf{k_\perp}\cdot \mathbf{r}} \notag\\
   && \frac{\phi_{n_1,q_1}(Z)\phi_{n_2,q_2}(Z)
\phi_{n_1,q_1}^*(Z^\prime)\phi_{n_2,q_2}^*(Z^\prime)}
  {E-\epsilon_{n_1}(q_1)-\epsilon_{n_2}(q_2)-\hbar^2k_\perp^2/m},
  \label{GEexplicit}
\end{eqnarray}
 where $q_B=\pi/d$ is the Bragg wave-vector and
$\phi_{n,q}(z_1)$ are the eigenstates of 
$H_z=-(\hbar^2/2m)d^2/dz_1^2+V_{opt}(z_1)$ with energies $\epsilon_n(q)$. 
In the following the periodic
potential will be taken of the form 
\be\label{vopt}
V_{opt}(z_1)=s E_R \sin^2\left(\frac{\pi z_1}{d}\right),
\ee
as usually provided by two counterpropagating laser beams (optical lattices),
where $s$ is the laser intensity, the spacing $d$ being related to the
wavelength of the laser fields  and $E_R=\hbar^2 \pi^2 /2m d^2$ is the recoil energy.
Since the singular term in the rhs of Eq.(\ref{defBC}) does not depend
on the external potential, we add and substract from $G_E$,
the Green function (\ref{GEexplicit}) evaluated for $s=0$. This 
permits us to write the kernel as
$K_E(Z,Z^{\prime })=\lim_{r\rightarrow 0}[G_E(\mathbf{r},Z;\mathbf{0},Z^{\prime})-G_E^{0}(\mathbf{r},Z;\mathbf{0},Z^{\prime})]+K_E^{0}(Z,Z^\prime)$, with
$K_E^{0}$ defined in Eq.(\ref{GEfree}).
After these manipulations, the integration over $\mathbf{k}_\perp$ 
in Eq.(\ref{GEexplicit}) converges and can be performed analytically. 
The remaining summation over band indices and
the integration over quasi-momentum are done numerically. 
\begin{figure}[tb]
\begin{center}
\includegraphics[width=8cm,angle=0,clip]{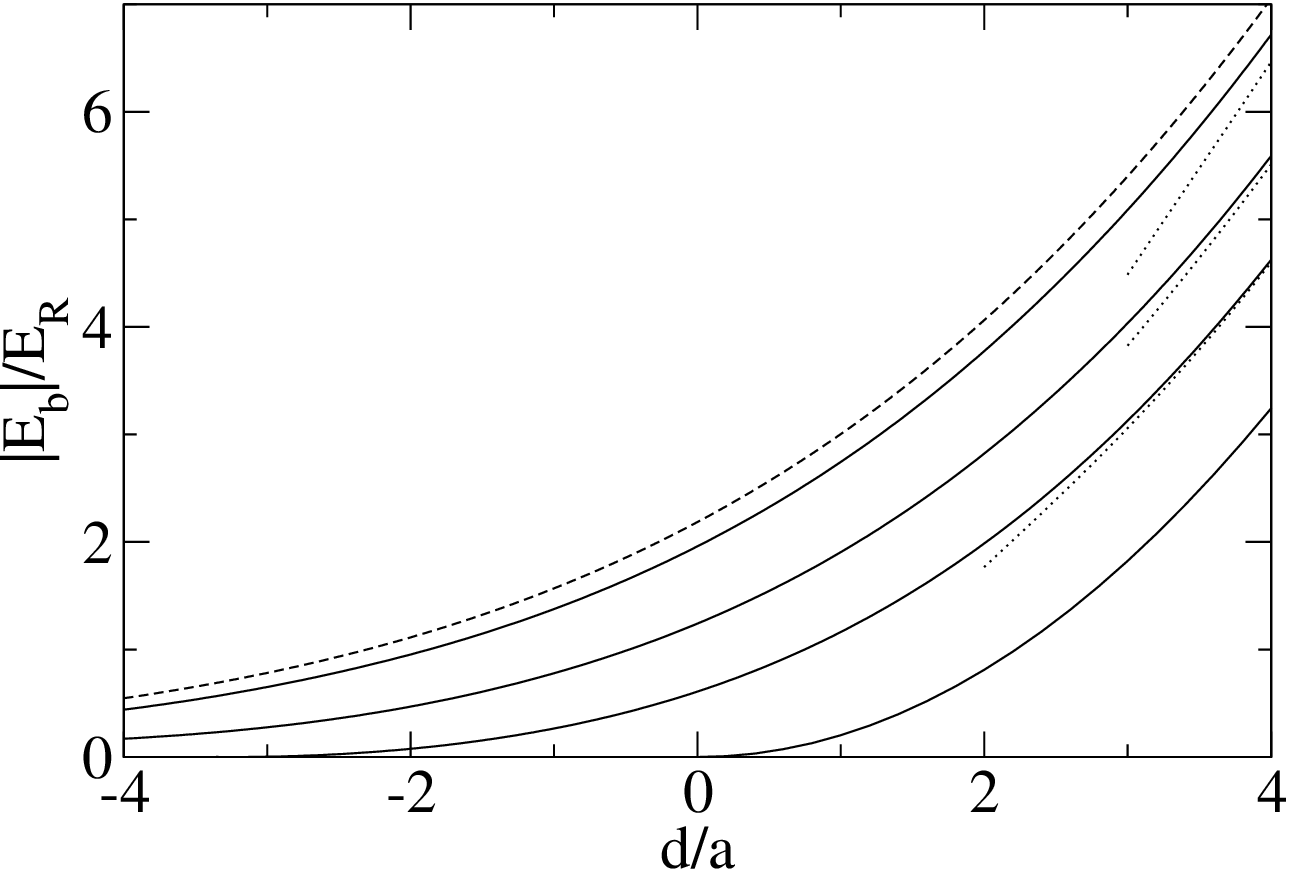}
\caption{Binding energy versus inverse scattering length for 
different values of the laser intensity: from top to bottom 
$s=20,10,5,0$ (solid line).
Also shown are the asymptotic behaviour for large $d/a$ 
(dotted line) and the binding energy in harmonic 
approximation (dashed line) for $s=20$.}
\label{figure1}
\end{center}
\end{figure}
We define the binding energy as $E_b(Q)=E(Q)-E_{ref}(Q)$,
where $E_{ref}(Q)=2\epsilon_1(Q/2)$ is the lowest energy state for two 
non interacting particles with total quasi-momentum $Q$ and $\epsilon_1$ 
is the dispersion of the lowest Bloch band. 
With this notation, $E_b$ is always negative for a bound state.

The binding energy for quasimomentum $Q=0$ as a function of the inverse 
scattering length is shown in Fig.\ref{figure1} 
for different values of the laser intensity. 

When the binding energy is large with respect to the depth of the lattice
$(h^2/ma^2 \gg sE_R)$, the periodic potential (\ref{vopt}) can be 
treated as a perturbation and
the wavefunction takes the form $\Psi(\mathbf r,Z)=\varphi (Z) \phi_b(r)$, 
with  $\phi_b(r)=(1/\sqrt{2\pi a}) e^{-r/a}/r$. 
Within first order perturbation 
theory, the effective potential acting on the center of mass of the molecule 
is given by
$U_{eff}(Z)=\int \phi_b^2(r) V(Z,z)d^3 \mathbf r$,
yielding
\be
\label{pot-ef}
U_{eff}(Z)=sE_R-sE_R\frac{2}{aq_B}\arctan(\frac{aq_B}{2})\cos(2q_BZ).
\ee
\begin{figure}[tb]
\begin{center}
\includegraphics[width=8cm,angle=0,clip]{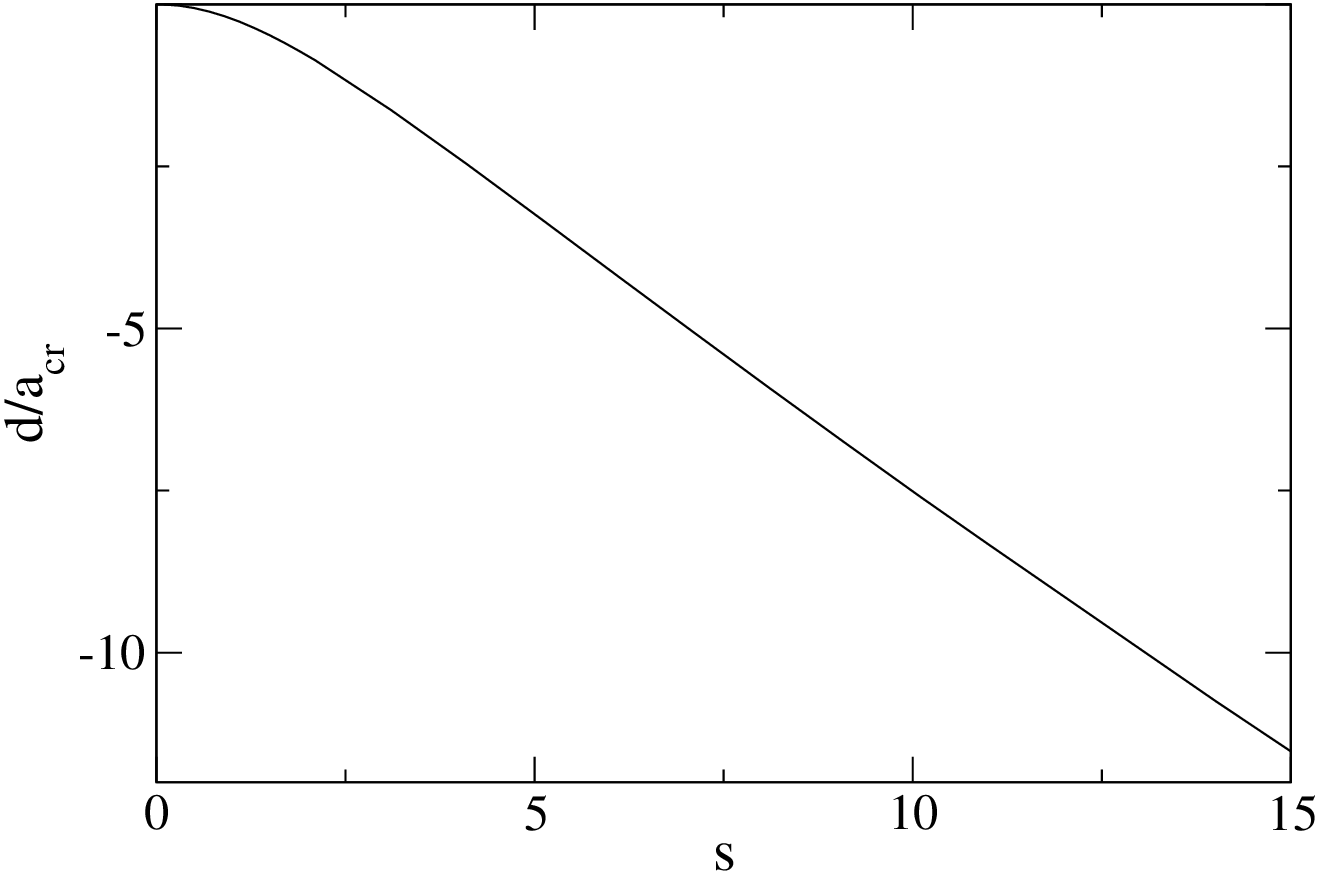}
\caption{ Critical value of the 
inverse scattering length as a function of the laser intensity.}
\label{figure2}
\end{center}
\end{figure} 
We see from Eq.(\ref{pot-ef}) that the depth of the effective potential 
depends on the scattering length. In the limit $a\ll d$, the external 
potential is constant on the scale of the size of the molecule and one finds 
$U_{eff}(Z)=2V_{opt}(Z)$. When $a$ increases, the size of the molecule
becomes comparable to the lattice spacing and the effective potential 
(\ref{pot-ef}) is reduced.
For $h^2/ma^2 \gg sE_R$, the binding energy is given by 
\be\label{eb}
E_b(Q)=-\hbar^2/m a^2+E'(Q)-E_{ref}(Q),
\ee
 where
 $E'$ can be found by solving the Schrodinger equation
\be\label{deep}
 \Big[-\frac{\hbar^2}{4m}\nabla_z^2+U_{eff}(Z)\Big]\varphi(Z)=E'\varphi(Z).  
\ee 
The asymptotic behaviour based in Eqs (\ref{pot-ef})-(\ref{deep}) 
is plotted in Fig.\ref{figure1} with dotted line
for different value of $s$. 

When the laser intensity becomes large $(s\gg 1)$
 and the binding energy is large compared to the 
width of the lowest Bloch band, the system
enters the quasi-2D regime. In this limit, the two interacting
atoms are localized at the bottom of the same optical well 
where, to a first approximation,  the potential (\ref{vopt}) 
is harmonic with frequency $\omega_{0}=2\sqrt{s}E_R/\hbar$.
In this case, the center of mass and the relative motion 
decouple and the scattering problem can be solved analytically 
\cite{gora,sbiscek}. The corresponding result for the binding energy
as a function of the inverse scattering length 
is shown in Fig.\ref{figure1} for $s=20$ (dashed line).
\begin{figure}[tb]
\begin{center}
\includegraphics[width=8cm,angle=0,clip]{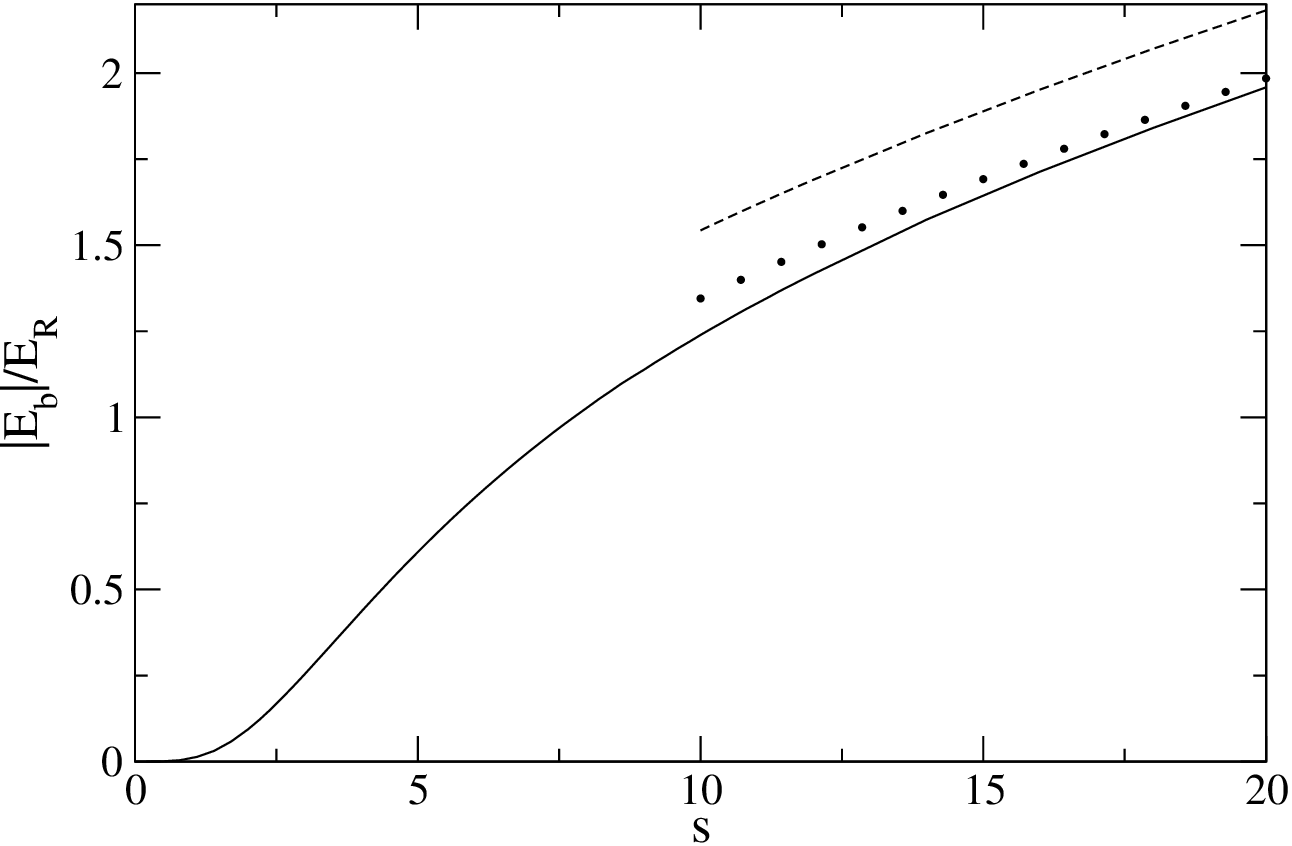}
\caption{Binding energy at unitarity $(d/a=0)$ 
as a function of the laser intensity (solid line). Also shown are
the harmonic approximation (dashed line) and the inclusion of 
anharmonic corrections (solid circles).}
\label{figure3}
\end{center}
\end{figure}
We see that the harmonic approximation slightly overestimates the 
correct binding energy. 

An important consequence of the presence of the optical lattice is the 
formation of a
bound state also for negative scattering length. However,
unlike in the asymptotic 2D limit,
the bound state disappears for values of the inverse scattering
length below a critical value $d/a_{cr}$. The dependence of $d/a_{cr}$
as a function of the laser intensity is shown in Fig.\ref{figure2}.
 We see that $d/a_{cr}=0 $
for $s=0$ and $d/a_{cr}\rightarrow -\infty$ as $s\rightarrow \infty$.
  
The behaviour of the binding energy as a function of the scattering length 
sufficiently close to the critical point is of the form
$1/a-1/a_{cr}=\sqrt{|E_b|m^*}/ \hbar C$, where $m^*$ is the atomic effective
mass evaluated at the bottom of the lowest Bloch band and $C=C(s)$ is a 
dimensionless function of the laser intensity.
For $s=0$ one has $C=1$ while 
in tight binding limit one finds $C=d/\sqrt{2 \pi}\sigma$, where
$\sigma$ is the width of the variational gaussian ansatz 
$w(z)=\exp(-z^2/2\sigma^2)/\pi^{1/4}\sigma^{1/2}$ for the  
Bloch function $\phi_{1,q}(z)\sim \sum_{\ell}e^{i q\ell d} w(z-\ell d)$.
This is just the renormalization factor
of the coupling constant $g$ obtained by integration over the gaussian 
profile \cite{pedri}. The above results show that the optical lattice gives 
rise to an effective shift of the resonance, in analogy with the case of
a 3D tight optical lattice investigated in Ref.\cite{Fedichev_PRL92}. 

It is interesting to discuss explicitly the behaviour of the binding
energy at resonance $(1/a=0)$ as a function of the laser intensity.
The numerical result is shown in Fig.\ref{figure3} (solid line).
 For large $s$  
the harmonic approximation gives
$E_b^{ho}(1/a=0)=-0.244 \hbar \omega_{0}=-0.488\sqrt{s}E_R$
(dashed line). The remaining shift $\Delta E$ in the binding 
is due to the non harmonic terms. 
We have found that the inclusion of such corrections via first order
perturbation theory results in 
an $s$-independent shift $\Delta E=0.198 E_R$ which well 
reproduces the full numerical result for $s\gtrsim 10$ 
(see dots in Fig.\ref{figure3}).

\begin{figure}[tb]
\begin{center}
\includegraphics[height=5.5cm,angle=0,clip]{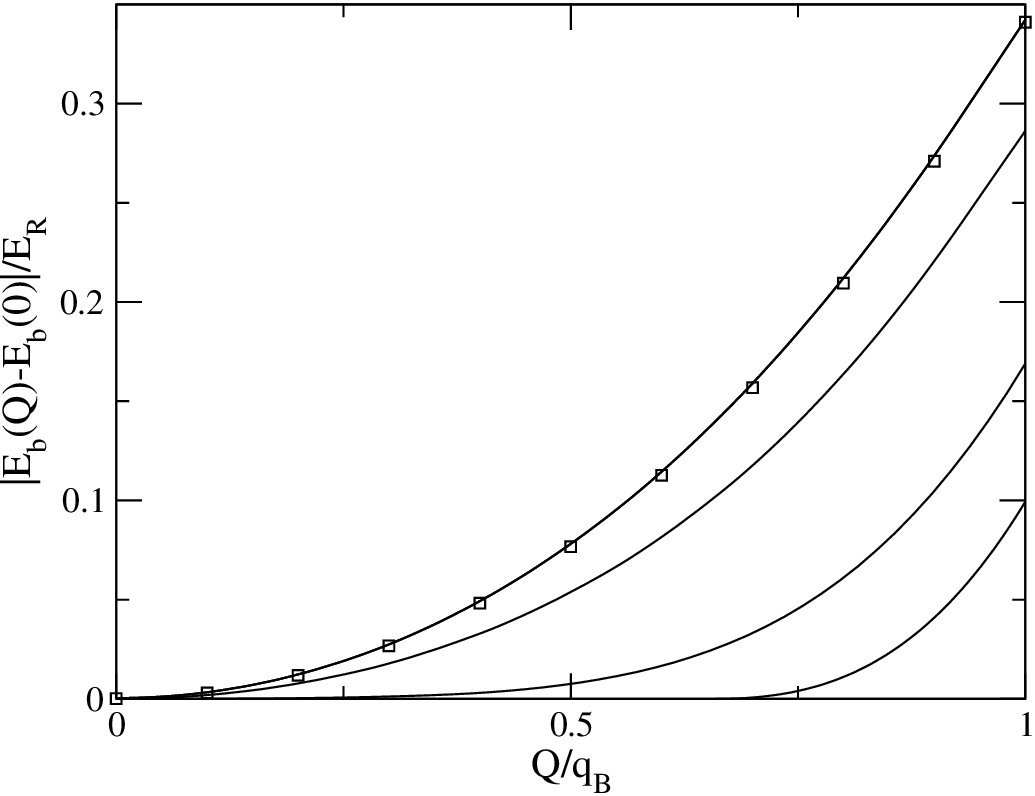}
\caption{Binding energy dispersion as a function of quasi-momentum
$Q$ for fixed $s=2.5$ and for different values of the inverse scattering length: from
top to bottom $d/a=2,0,-1.17(= d/a_{cr}),-1.66$. 
The solid squares give the dispersion at $d/a=2$ evaluated from 
Eqs (\ref{pot-ef})-(\ref{deep}).} 
\label{figure4}
\end{center}
\end{figure}

So far we have discussed the results for the binding energy
corresponding to vanishing values of the quasi-momentum.
Differently from the case of harmonic trapping, where the binding energy
does not depend on the center of mass motion,
in the presence of an optical lattice the
binding energy of the molecule depends on its quasi-momentum.   
This is plotted in Fig.\ref{figure4} 
for $s=2.5$ and different value of the scattering length. 
We see that the modulus of the binding energy increases when $Q$
increases. Moreover,
when the scattering length is negative and $d/a$
crosses the critical value $d/a_{cr}$, the bound state 
with zero quasi-momentum breaks first. 
\begin{figure}[t]
\begin{center}
\includegraphics[width=8cm,angle=0,clip]{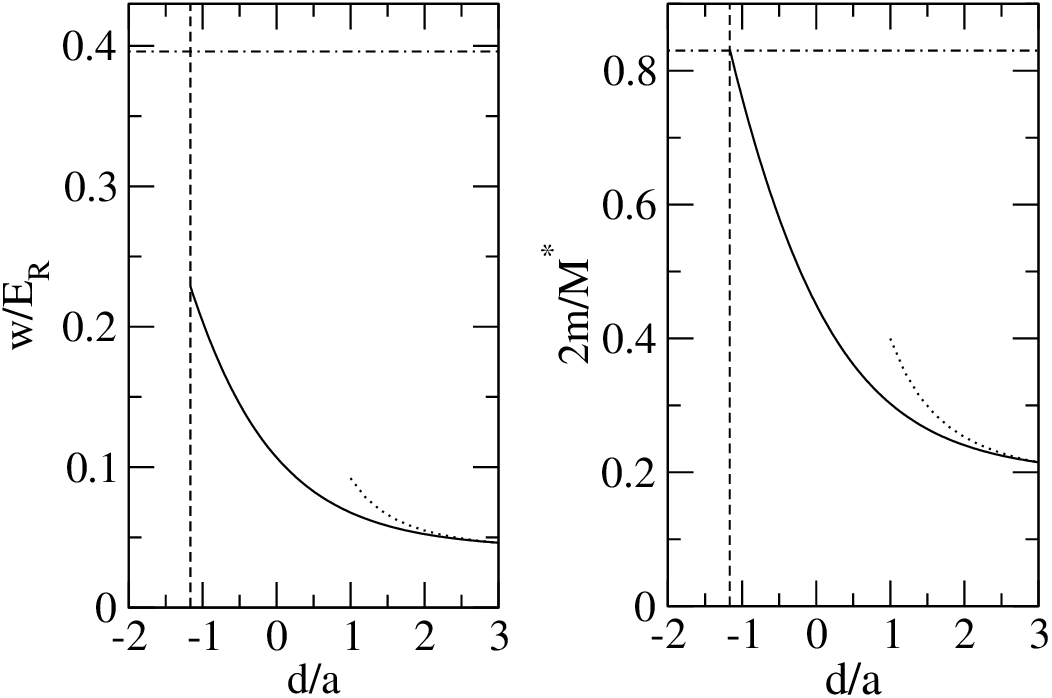}
\caption{ Bandwidth ({\sl Left Panel}) and effective mass ratio 
$2m/M^*$ ({\sl Right Panel}) of the molecule as a function of the inverse scattering length
for $s=2.5$ (solid line). 
The bandwidth at the critical point is always smaller  
than $2\epsilon_1(q_B/2)$ (dot-dashed line) [See text].
Conversely, at the critical point the ratio $2m/M^*$ coincides with
the effective mass ratio $m/m^*$ for single atoms (dot-dashed line).
The asymptotic behaviour for large positive $d/a$ 
is also shown (dotted line). }
\label{figure5}
\end{center}
\end{figure}
The bandwidth, defined as $w=E(q_B)-E(0)$, is plotted in 
Fig.\ref{figure5} {\sl (Left Panel)}  as a function of $d/a$ for $s=2.5$.
For a given value of the laser intensity, the curve stops at $d/a=d/a_{cr}$ 
where 
$w=2\epsilon_1(q_B/2)+E_b(q_B)<2\epsilon_1(q_B/2)$ because the binding 
energy at $Q=q_B$ is finite at the critical point (see Fig.\ref{figure4}). 
The bandwidth for molecules is always smaller than the one for single atoms.
For instance for $s=2.5$ the width of the lowest band for atoms is $0.52E_R$. 
This effect results from the correlated motion of the two constituent
atoms. This is best seen for $h^2/ma^2 \gg sE_R$, where 
the depth of the effective potential (\ref{pot-ef}) felt by the 
molecule increases as $d/a$ increases. 
In the limit $d/a \gg 1$, $U_{eff}(Z)=4sE_{Rm}\cos^2(\pi Z/d)$, 
where $E_{Rm}=E_R/2$ is the recoil energy for a particle with mass $2m$, 
yielding an effective laser intensity $s_m=4s$. 
In particular, in tight binding approximation the atomic bandwidth 
is proportional to the WKB tunneling exponent $e^{-S_0}$,
where $S_0=\pi^2s^{1/2}/4$, while the bandwidth for molecules is
much smaller, being proportional to $e^{-2S_0}$. 

In Fig.\ref{figure5} {\sl (Right Panel)} we show the results for 
the effective mass of the molecule, defined through 
$1/M^*=(\partial^2 E(Q)/\partial Q^2)_{|Q=0}$, 
as a function of $d/a$ for laser intensity $s=2.5$.
We see that, in complete analogy with the bandwidth, the
ratio $2m/M^*$ decreases as $d/a$ increases.  
At the critical point where the molecule breaks, one finds
$M^*=2m^*$, where $m^*$ is the effective mass of a single atom.

In the last part of the Letter we discuss the behaviour of the size
of the molecule as a function of the scattering length and the laser 
intensity.
The typical size of the molecule in the radial direction
is fixed by the binding energy according to 
$\lambda_\perp\approx \hbar/\sqrt{|E_b|m}$, where $E_b$ is fixed by 
the scattering length [see Fig.(\ref{figure1})].
Due to the optical confinement, the molecule is more squeezed in the
axial direction, so generally $\lambda_z \leq \lambda_\perp$.
For large $d/a$ the molecule is not distorted by the optical lattice and
therefore $\lambda_\perp \approx \lambda_z \approx a$. 
When $d/a$ decreases, the wavefunction becomes more elongated
in the radial direction. As long as the binding energy is large 
compared to the width of the lowest Bloch band, the two atoms 
stay in the same optical well and therefore $\lambda_z \lesssim d$. 
When this condition is violated,
the two atoms can hop to different lattice sites and the wavefunction becomes 
spread also in the $z$ direction. Finally, close to the critical point, 
one has $\lambda_\perp,\lambda_z \gg d$ with the 
anisotropy ratio fixed by the simple relation 
$\lambda_\perp/\lambda_z= \sqrt{m^*/m}$.

In conclusion, we have presented an exact numerical method to calculate the
binding energy and the tunneling properties of a molecule
  formed by two atoms interacting via a short range potential,
 in the presence of a 1D optical lattice, for {\sl arbitrary}
 values of laser intensity, scattering length and
 quasi-momentum of the molecule. The theoretical predictions could be
checked in experiments on dilute ultracold gases. 
They could also have important consequences on the
many-body properties of these systems.  In particular the formation of
molecules induced by the periodic optical field is expected to modify the
behaviour of the BEC-BCS crossover in a degenerate Fermi gas.

We acknowledge interesting discussions with Z. Idziaszek and C. Salomon.
This work was supported by the Ministero dell'Istruzione, 
dell'Universita' e della Ricerca (M.I.U.R.), by 
the Special Research Fund of the University of Antwerp, BOF NOI UA 2004, 
and by the FWO-V project No.G.0435.03. M.W. is financially 
supported by the `FWO -- Vlaanderen'.

\end{document}